\documentclass[preprint,12pt]{elsarticle}

\usepackage{amsmath,amssymb,graphicx}
\usepackage{multirow}
\usepackage[colorlinks]{hyperref}

\journal{Physics Letters A}

\begin{document}

\begin{frontmatter}



\title{Exact quantum search based on analytical multiphase matching for
known number of target items and the experimental demonstration on
IBM Q}


\author{Tan Li$^{a,b}$, Xiang-Qun Fu$^{a,b}$, Yang Wang$^{a,b}$, Shuo Zhang$^{a,b}$, Xiang
Wang$^{a,b}$, Yu-Tao Du$^{a,b}$, and Wan-Su Bao$^{a,b}$\corref{cor1}}

\cortext[cor1]{Corresponding author at: SSF IEU, Zhengzhou 450001, China. E-mail address: bws@qiclab.cn (W. -S. Bao).}

\address{$^a$ Henan Key Laboratory of Quantum Information and Cryptography, SSF IEU, Zhengzhou, Henan 450001, China\\
$^b$ Synergetic Innovation Center of Quantum Information and Quantum Physics, University of Science and Technology of China, Hefei, Anhui 230026, China}

\begin{abstract}
In [Phys. Rev. Lett. 113, 210501 (2014)], to achieve the optimal fixed-point quantum search in the case of unknown fraction (denoted by $\lambda$) of target items, the analytical multiphase matching (AMPM) condition has been   proposed. In this paper, we find out that the AMPM condition can also be used to design the exact quantum search algorithm in the case of known $\lambda$, and the minimum number of iterations reaches the optimal level of existing exact algorithms. Experiments are performed to demonstrate the proposed algorithm on IBM's quantum computer. In addition, we theoretically find two coincidental phases with equal absolute value in our algorithm based on the AMPM condition and that algorithm based on single-phase matching. Our work confirms the practicability of the AMPM condition in the case of known $\lambda$, and is helpful to understand the mechanism of this condition.
\end{abstract}

\begin{keyword}


Quantum search \sep
Analytical multiphase matching \sep
Known number of target items \sep
100\% success probability \sep
IBM quantum experience

\end{keyword}

\end{frontmatter}


\section{Introduction \label{sec:Introduction}}

Grover quantum search algorithm \cite{Grover1996,Grover1997} is one
of the great quantum algorithms, which achieves a quadratic speedup
over classical search algorithms. Many generalizations and variants
of Grover's algorithm have been studied \cite{Grover1998,Boyer1998,Biron1999,Hyer2000,Younes2004,Grover2005,Li2014,Dalzell2017,Byrnes2018,Li2018,Toyama2019}, especially the multiphase
matching methods.

In 2008, Toyama et
al. found the original multiphase matching condition \cite{Toyama2008}
by means of numerical fitting, i.e.,
\begin{equation}
\phi_{j}=\varphi_{l+1-j},
\end{equation}
to achieve success probability close to 100\% over a wide
range of $\lambda$ through a small number of iterations $\prod_{j=1}^{l}G\left(\phi_{j},\varphi_{j}\right)$, where $G$ is the generalized Grover operation defined by Eq.~(\ref{eq:arbitrary-phase-G-operation}).
However, the whole range of $\lambda\in(0,1)$ cannot be covered \cite{Toyama2008,Toyama2009}.

Fortunately, in 2014, Yoder et al. proposed the analytical multiphase
matching (AMPM) condition \cite{Yoder2014}
which gives the analytical forms of phases $\left\{ \phi_{j},\varphi_{j}\right\}$ (see Eq.~(\ref{eq:multiphase-matching-condition}) for details),
and obtained an algorithm which can
always maintain a success probability no less than $1-\delta^{2}$
on the range of $\lambda\in\left[\lambda_{0},1\right)$, where $\delta\in\left(0,1\right]$ and
$\lambda_{0}$ is an arbitrary lower bound of $\lambda$.
This algorithm also possesses the fixed-point property and achieves
a quadratic speedup over the original fixed-point quantum search algorithm
\cite{Grover2005},
showing the great fascination of the AMPM condition.

However, Ref.~\cite{Yoder2014} only aims to deal with the case of unknown $\lambda$. Then,
in the case of known $\lambda$, what kind of performance will the AMPM condition have?
In this paper, we expect to design a quantum search algorithm with 100\% success probability for the known $\lambda$ based on the AMPM condition and implement a proof-of-principle experiment, to reveal more applications of this condition.

The paper is organized as follows. Section~\ref{sec:Exact_MPM} describes
our analytical matched-multiphase quantum search
algorithm with 100\% success probability. Section~\ref{sec:Experimental-implementation}
introduces the experimental implementation of the proposed algorithm
on IBM's cloud quantum computing platform. Section~\ref{sec:Discussions} discusses
the connection between the AMPM condition
and the single-phase matching condition, followed by a brief conclusion
in Section~\ref{sec:Conclusion}.

\section{Exact quantum search based on the analytical multiphase matching condition
\label{sec:Exact_MPM}}
\begin{figure}[b]
\begin{centering}
\includegraphics[width=13.5cm]{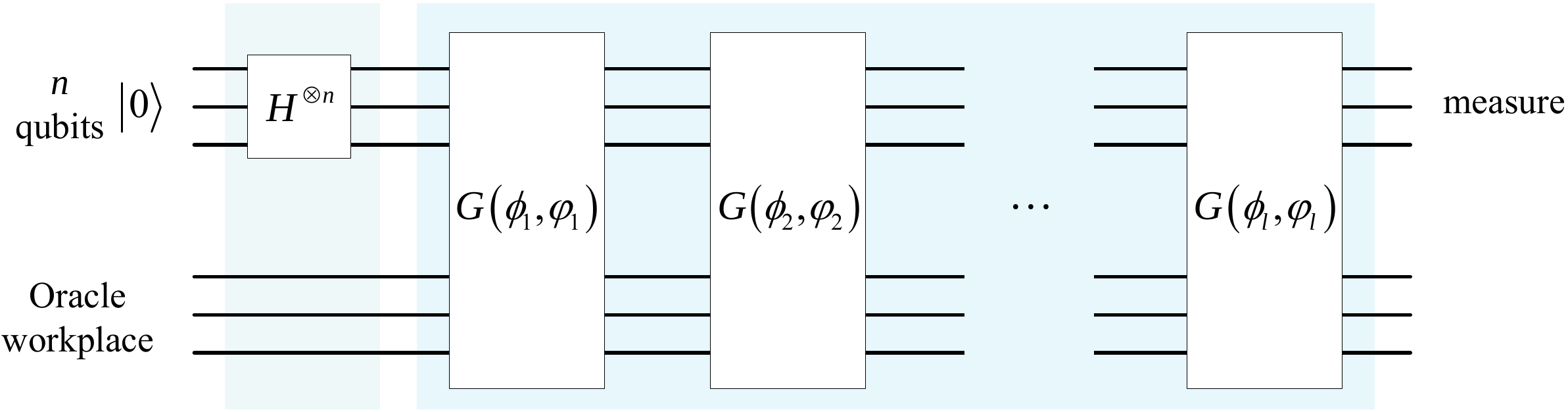}
\par\end{centering}
\protect\caption{(Color online.) Schematic circuit for the exact matched-multiphase quantum search
algorithm. \label{fig:schematic-circuit-algorithm}}
\end{figure}

In order to characterize the performance of the AMPM condition \cite{Yoder2014} in case of known $\lambda$,
we expect to
design a quantum search algorithm with 100\% success probability.
The specific steps are described as follows (the corresponding flow
diagram is shown in Fig.~\ref{fig:schematic-circuit-algorithm}).

Step 1: Prepare the initial state $\left|\psi\right\rangle =H^{\otimes n}\left|0\right\rangle $,
which can be written as
\begin{equation}
\left|\psi\right\rangle =\sqrt{\lambda}\left|\alpha\right\rangle +\sqrt{1-\lambda}\left|\beta\right\rangle ,
\end{equation}
where $H$ is the Hadamard transform,
$\left|\alpha\right\rangle $
($\left|\beta\right\rangle $) is the equal superposition of all target
(nontarget) states, i.e.,
\begin{eqnarray}
\left|\alpha\right\rangle  & = & \frac{1}{\sqrt{M}}\sum\limits _{x\in f^{-1}\left(1\right)}\left|x\right\rangle,\label{eq:target_state}\\
\left|\beta\right\rangle  & = & \frac{1}{\sqrt{N-M}}\sum\limits _{x\in f^{-1}\left(0\right)}\left|x\right\rangle,\label{eq:nontarget_state}
\end{eqnarray}
and $\lambda=M/N$, $M$ is the
number of target items in the database of size $N$.

Step 2: Perform on $\left|\psi\right\rangle $ the sequence of matched-multiphase
Grover operations $\prod_{j=1}^{l}G\left(\phi_{j},\varphi_{j}\right)$, where
$G\left(\phi,\varphi\right)$ is the Grover iteration with
arbitrary phases, i.e.,
\begin{eqnarray}
G\left(\phi,\varphi\right) & = & -HS_{0}^{\phi}HS_{f}^{\varphi},\label{eq:arbitrary-phase-G-operation}
\end{eqnarray}
$S_{f}^{\varphi}$ and $S_{0}^{\phi}$ represent the selective phase shifts,
\begin{eqnarray}
S_{f}^{\varphi} & = & I-\left(-e^{i\varphi}+1\right)\sum_{x\in f^{-1}\left(1\right)}\left|x\right\rangle \left\langle x\right|,\\
S_{0}^{\phi} & = & I-\left(-e^{i\phi}+1\right)\left|0\right\rangle \left\langle 0\right|,
\end{eqnarray}
$l$ is the number of iterations, satisfying
\begin{equation}
l\ge\left\lceil \frac{\pi}{4\arcsin\sqrt{\lambda}}-\frac{1}{2}\right\rceil \equiv l_{{\rm min}},\label{eq:l-condition}
\end{equation}
and $\left\{ \phi_{j},\varphi_{j}\right\} $ meet the analytical
multiphase matching condition \cite{Yoder2014}, i.e.,
\begin{equation}
\phi_{j}=\varphi_{l-j+1}=-2{\rm arccot}\left(\sqrt{1-\gamma^{2}}\tan\left(2\pi j/L\right)\right),\thinspace{\rm for}\thinspace1\le j\le l,\label{eq:multiphase-matching-condition}
\end{equation}
where $L=2l+1$, $\gamma=T_{1/L}^{-1}\left(1/\delta\right)$, $T_{L}\left(x\right)=\cos\left[L\arccos\left(x\right)\right]$
is the $L^{{\rm th}}$ Chebyshev polynomial of the first kind \cite{Mason2002},
and
\begin{equation}
\delta=T_{L}^{-1}\left(\frac{\cos\left(\frac{\pi}{2L}\right)}{\sqrt{1-\lambda}}\right).\label{eq:delta-condition}
\end{equation}

Step 3: Measure the system. This will produce one of the marked states
with 100\% success probability.

Reasons for the selection of $l$ of Eq.~(\ref{eq:l-condition})
and $\delta$ of Eq.~(\ref{eq:delta-condition}) are as follows.
First, according to Ref.~\cite{Yoder2014}, the final state of Step 2 can be written as
\begin{equation}
\left|C_{L}\right\rangle =\sqrt{P_{L}}\left|\alpha\right\rangle +\sqrt{1-P_{L}}\left|\beta\right\rangle ,\label{eq:final-state-C_L}
\end{equation}
where $P_{L}$ denotes the success probability,
\begin{equation}
P_{L}=1-\delta^{2}T_{L}^{2}\left[T_{1/L}\left(1/\delta\right)\sqrt{1-\lambda}\right],\label{eq:success_probability}
\end{equation}
Then, let $P_{L}=1$, we can obtain all the $k$ maximum points of $P_{L}$,
denoted as
\begin{equation}
\lambda_{l,j}^{\delta,\max}=1-\gamma^{2}\cos^{2}\left(\frac{2j-1}{2L}\pi\right),\thinspace1\le j\le l,
\end{equation}
and \begin{eqnarray}
\min\left\{ \lambda_{l,j}^{\delta,\max}:1\le j\le k\right\}  & = & \lambda_{l,j=1}^{\delta,\max},\\
\min\left\{ \lambda_{l,j=1}^{\delta,\max}:0<\delta\le1\right\}  & = & \lambda_{l,j=1}^{\delta=1,\max}.
\end{eqnarray}
Note that, making the success probability at $\lambda$ reach 100\%
is the equivalent of
having $\lambda$ happen to be a certain maximum point,
and the range of maximum points $\big\{\lambda_{l,j}^{\delta,\max}\big\}$ is $\big[\lambda_{l,j=1}^{\delta=1,\max},1\big)$.
Therefore, $\lambda\ge\lambda_{l,j=1}^{\delta=1,\max}$ is required, yielding the result of Eq.~(\ref{eq:l-condition}) of the number
of iterations $l$.
Let $\lambda=\lambda_{l,j=1}^{\delta,\max}$, we can further obtain
the condition Eq.~(\ref{eq:delta-condition}) of $\delta$.
Finally,
the phases $\left\{ \phi_{j},\varphi_{j}\right\} $ can be
calculated from Eq.~(\ref{eq:multiphase-matching-condition}).

\begin{figure}[t]
\begin{centering}
\includegraphics[width=8cm]{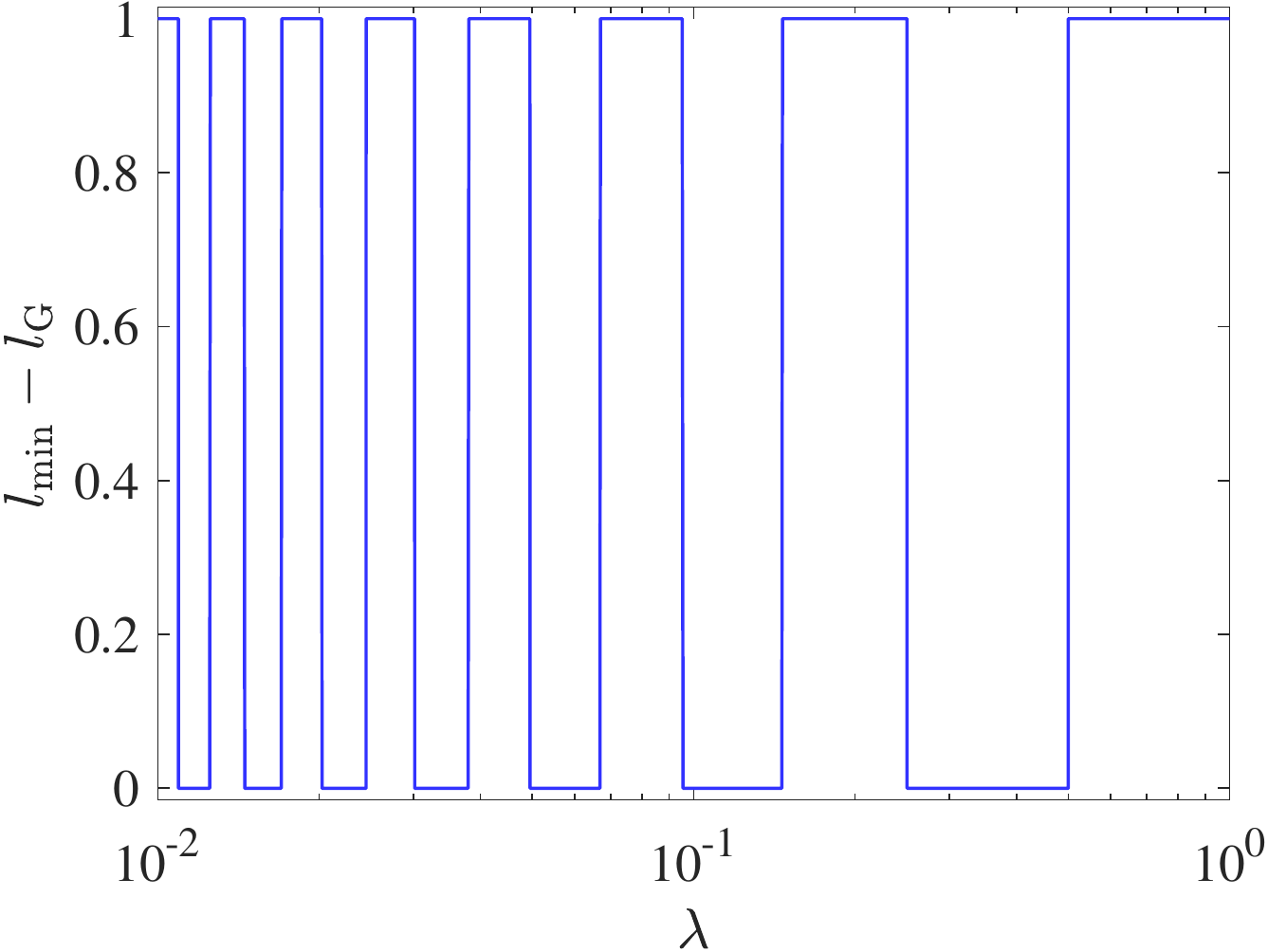}
\par\end{centering}
\protect\caption{(Color online.) Difference between the minimum number of iterations ($l_{{\rm min}}$) of our algorithm and the optimal number of iterations ($l_{{\rm G}}$)
of the original Grover algorithm as a function of the fraction
$\lambda$ of target items. \label{fig:iterations-diff-vs-lambda}}
\end{figure}
At this point, all the selection methods of parameters in our exact algorithm have been obtained.
Figure~\ref{fig:iterations-diff-vs-lambda}
plots the minimum number of iterations ($l_{{\rm min}}$) of our algorithm
and the optimal number of iterations ($\left\lceil \frac{\pi}{4\arcsin\sqrt{\lambda}}\right\rceil -1\equiv l_{{\rm G}}$ \cite{Nielson2000})
of the original Grover algorithm
as a function of $\lambda$, which shows that the difference is up to one.
In fact, it can be found that our minimum number of iterations $l_{{\rm min}}$ of Eq.~(\ref{eq:l-condition})
is just the same as that of the existing exact quantum search algorithms \cite{Long2001a,Toyama2013,Hu2002,Hsieh2004,Brassard2002},
according to Eq.~(7) of Ref.~\cite{Long2001a}, Eqs.~(15), (20) and (23)
of Ref.~\cite{Hsieh2004}, and Theorem 4 of Ref.~\cite{Brassard2002}.

\section{Experimental implementation \label{sec:Experimental-implementation}
}

We also conducted an experiment to demonstrate the proposed algorithm
on the IBM's 5-qubit computer (ibmqx4) \cite{IBM}.

\begin{figure}[t]
\centering{}\includegraphics[width=13.7cm]{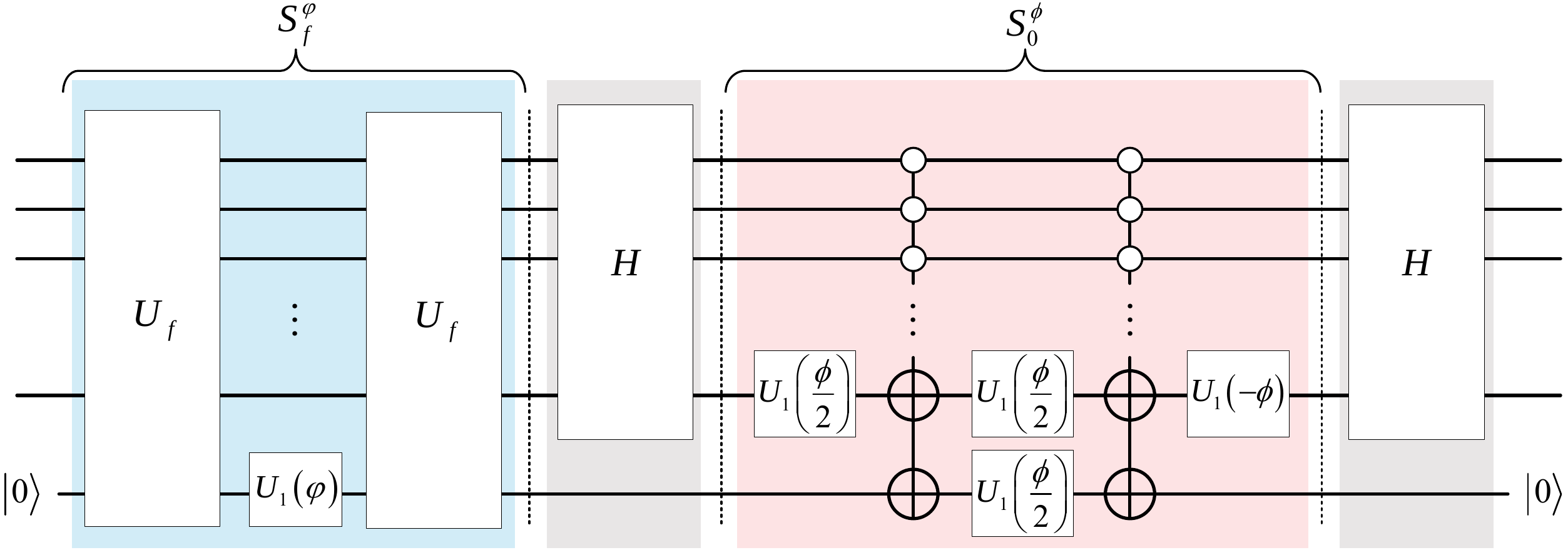}\protect\caption{(Color online.) Circuit for performing the generalized Grover operation $G\left(\phi,\varphi\right)$
of Eq.~(\ref{eq:arbitrary-phase-G-operation}). \label{fig:circuit-generalized-Grover-operation}}
\end{figure}

As shown in Fig.~\ref{fig:schematic-circuit-algorithm},
our algorithm is mainly composed of the generalized Grover operation
$G\left(\phi,\varphi\right)$, defined by Eq.~(\ref{eq:arbitrary-phase-G-operation}).
According to Ref.~\cite{Yoder2014}, a quantum circuit for realizing
$G\left(\phi,\varphi\right)$ is shown in Fig.~\ref{fig:circuit-generalized-Grover-operation},
where $U_{f}$ is the Oracle operator, $U_{f}\left|x\right\rangle \left|y\right\rangle =\left|x\right\rangle \left|y\oplus f\left(x\right)\right\rangle $,
$H$ is the Hadamard transform, and $U_{1}\left(x\right)$ is the quantum gate defined by
\begin{equation}
U_{1}\left(x\right)=\left[\begin{array}{cc}
1 & 0\\
0 & e^{ix}
\end{array}\right].
\end{equation}
Note that, for the single-qubit quantum search, $S_{0}^{\phi}$ can
be implemented simply by $U_{1}\left(-\phi\right)$, up to a global
phase.

\begin{figure}[b]
\begin{centering}
\includegraphics[width=7.8cm]{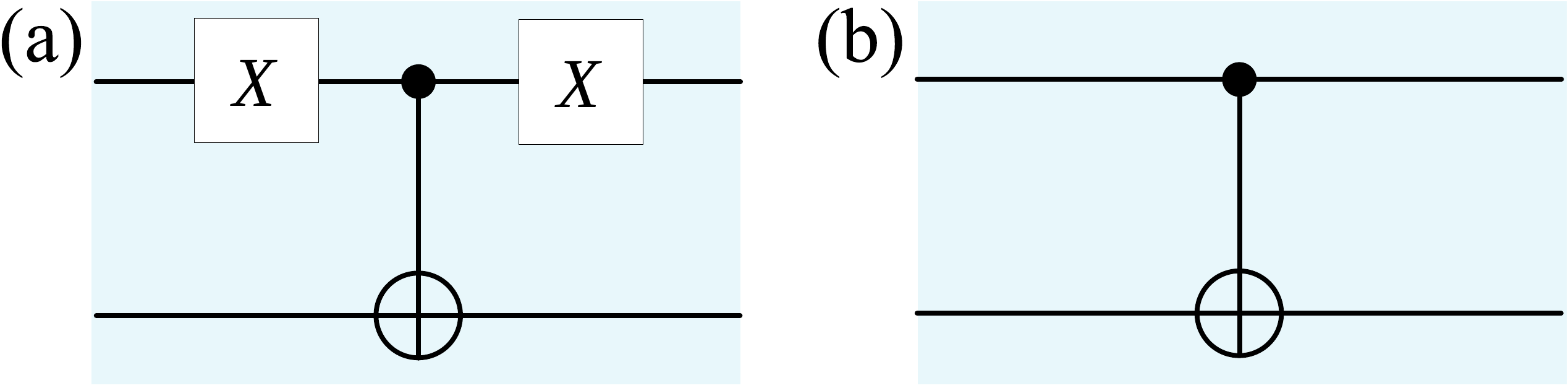}
\par\end{centering}
\protect\caption{(Color online.) Circuits for all the possible Oracles in the single-qubit and single-solution quantum search, where (a) $f\left(0\right)=1$ and (b) $f\left(1\right)=1$.
\label{fig:Oracles}}
\end{figure}

For the sake of simplicity, we chose the single-qubit ($N=2$) and single-solution ($M=1$)
example for experimentation. All the two possible Oracles are shown
in Fig.~\ref{fig:Oracles}.

\begin{figure}[t]
\begin{centering}
\includegraphics[width=13cm]{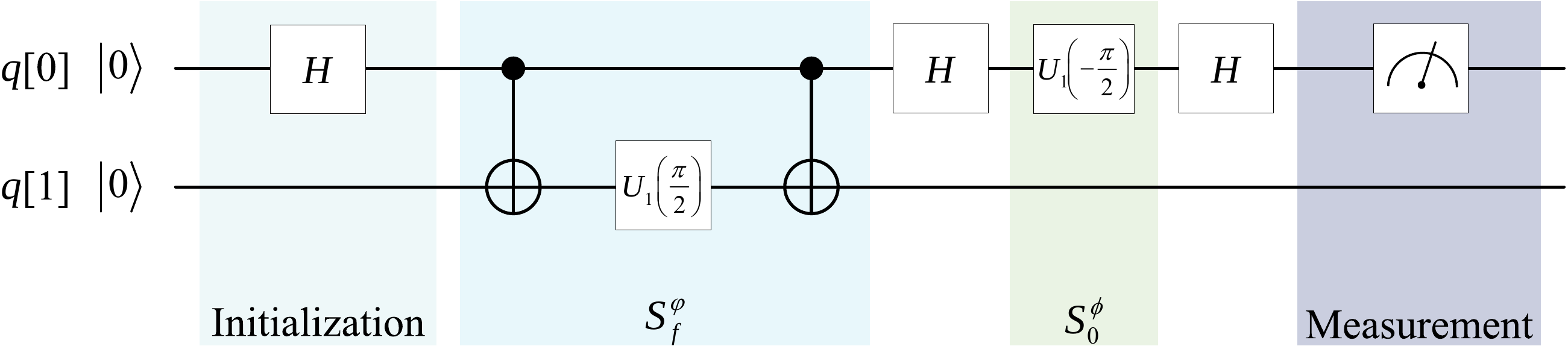}
\par\end{centering}
\protect\caption{(Color online.) Complete circuit for one iteration of the exact analytical matched-multiphase
quantum search with the Oracle corresponding to $f\left(1\right)=1$.
\label{fig:circuit-one-iteration}}
\end{figure}

From Eq.~(\ref{eq:l-condition}), it follows that the required number of iterations
$l\ge1$ for $\lambda=0.5$.
Here, we respectively choose $l=1$, $2$ and $3$
to verify the results of the
algorithm for the two possible Oracles.
Corresponding parameters are shown in Table~\ref{tab:values-multiple-phases}.
\begin{table}[t]
\centering{}%
\begin{tabular}{cccc}
\hline
\hline
 $l$ & $\delta$ & $\phi_{1},\phi_{2},\cdots,\phi_{l}$ & $\ensuremath{\varphi_{1},\varphi_{2},\cdots,\varphi_{l}}$\tabularnewline
\hline
1 & 0.272166 & $\ensuremath{\pi/2}$ & $\ensuremath{\pi/2}$\tabularnewline
\hline
\multirow{2}{*}{2} & \multirow{2}{*}{0.035103} & -0.904557 & 2.237036\tabularnewline
 &  & 2.237036 & -0.904557\tabularnewline
\hline
\multirow{3}{*}{3} & \multirow{3}{*}{0.005398} & -1.717287 & 2.501328\tabularnewline
 &  & 0.640265 & 0.640265\tabularnewline
 &  & 2.501328 & -1.717287\tabularnewline
\hline
\hline
\end{tabular}\protect\caption{Parameters of our exact algorithm for $\lambda=0.5$ with $l=1$, $2$ and $3$. \label{tab:values-multiple-phases}}
\end{table}
As an example, the complete
quantum circuit for $l=1$ with the Oracle corresponding to $f\left(1\right)=1$ is depicted in Fig.~\ref{fig:circuit-one-iteration}.

\begin{figure}[t]
\begin{centering}
\includegraphics[width=11.5cm]{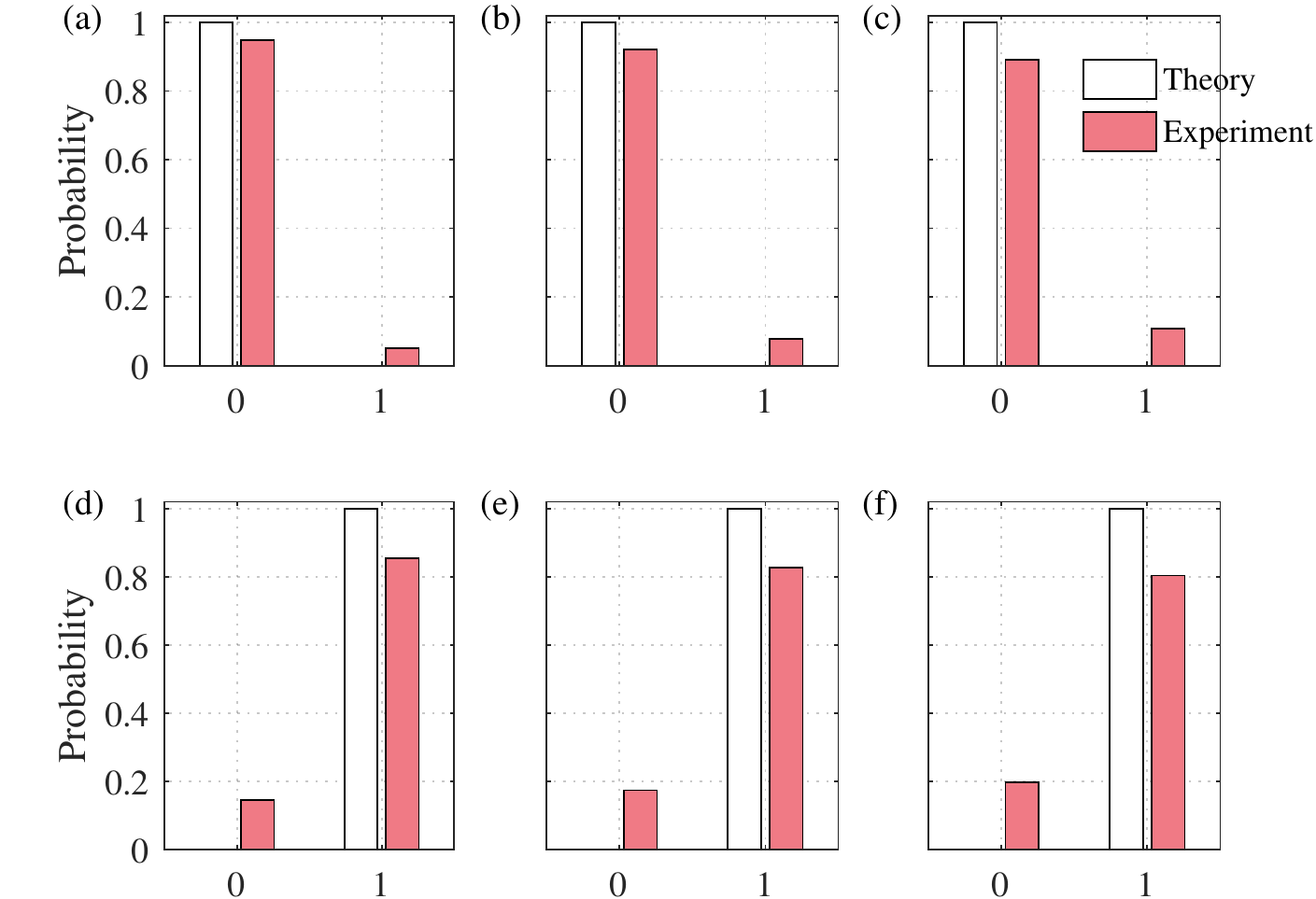}
\par\end{centering}
\protect\caption{(Color online.) Experimental results. (a-c) and (d-f) correspond to
$f\left(0\right)=1$ and $f\left(1\right)=1$, (a,d), (b,e) and (c,f) correspond to $l=1$, 2 and 3, and the white and red (grey)
bars correspond to the theoretical and experimental probabilities, respectively.\label{fig:experimental-results}}
\end{figure}

Figure~\ref{fig:experimental-results} exhibits the obtained experimental
results, where the white and red (grey) bars represent the theoretical
and experimental probabilities, respectively.
Ideally, when $f\left(0\right)=1$, the output is in $\left|0\right\rangle $
with a probability of 100\%, and another 0\% probability yields $\left|1\right\rangle $.
When $f\left(1\right)=1$, probabilities are just the
opposite. To characterize the overlap between experimental and theoretical
results, we adopt the statistical fidelity \cite{Huang2017,Huang2018}
\begin{equation}
F=\sum_{j=0}^{1}\sqrt{p_{j}^{{\rm exp}}p_{j}^{{\rm th}}},
\end{equation}
where $p_{j}^{{\rm th}}$ and $p_{j}^{{\rm exp}}$ represent the theoretical
and experimental output probabilities of the state $\left|j\right\rangle $, respectively.
According to the data in Fig.~\ref{fig:experimental-results}~(a-f),
the fidelities can be calculated as 0.9743, 0.9601, 0.9443, 0.9242, 0.9089
and 0.8966, which confirms the feasibility of our algorithm. The deviations
between experimental and theoretical results are mainly related
to errors in the readout and quantum gates. Calibration data provided
by the IBM platform shows that readout error is $3.60\times10^{-2}$
and gate error is $1.12\times10^{-3}$.

\section{Discussions \label{sec:Discussions}}

Although the AMPM condition \cite{Yoder2014}
differs significantly from the single-phase matching condition \cite{Long1999,Long2001,Long2002,Li2002},
we find out a coincidental connection between them. The specific discussions are
as follows.

In the exact quantum search algorithm \cite{Long2001a} based on the single-phase
matching condition \cite{Long1999}, the sequence of operations is
given by $G^l(\phi,\varphi)$, where $\phi$ and $\varphi$ meet the condition
\begin{equation}
\phi=\varphi,
\label{single-phase_matching_condition}
\end{equation}
and the value of phase $\phi$ is given by
\begin{equation}
\pm\arccos\left(1-\frac{1-\cos\left(\frac{\pi}{2l+1}\right)}{\lambda}\right)\equiv \phi_{{\rm s}}.
\label{eq:phi_s}
\end{equation}
When $\lambda=1/2$, $\phi_{{\rm s}}=\pm\frac{\pi}{2}$, $\pm0.904557$
and $\pm0.640265$ for $l=1$, 2 and 3. Compared with Table~\ref{tab:values-multiple-phases},
we can see that in the case of same number of iterations, it seems
that there is always a phase in the exact analytical multiphase matching quantum
search algorithm, of which the absolute value is exactly the same
as that of phase $\phi_{{\rm s}}$ in the exact single-phase matching algorithm.

In fact, we can prove that there indeed exists a phase $\phi_{m}$ in
$\left\{ \phi_{j}:1\le j\le l\right\} $, satisfying
\begin{equation}
\phi_{m}=\begin{cases}
\left|\phi_{{\rm s}}\right|, & {\rm if}\thinspace l\thinspace{\rm is\thinspace odd,}\\
-\left|\phi_{{\rm s}}\right|, & {\rm if}\thinspace l\thinspace{\rm is\thinspace even,}
\end{cases}\label{eq:relation-phi_m-phi_s}
\end{equation}
where
\begin{equation}                                                                                  m=\begin{cases}
\left(l+1\right)/2, & {\rm if}\thinspace l\thinspace{\rm is\thinspace odd,}\\
l/2, & {\rm if}\thinspace l\thinspace{\rm is\thinspace even.}
\end{cases}\label{eq:m}
\end{equation}

Reasons are as follows.
On the one hand, from Eq.~(\ref{eq:phi_s}) it follows that
\begin{equation}
\frac{\cot^{2}\left(\frac{\phi_{{\rm s}}}{2}\right)}{\lambda-\sin^{2}\left(\frac{\pi}{2L}\right)}=\frac{1}{\sin^{2}\left(\frac{\pi}{2L}\right)}.
\label{eq:phi_s_equ}
\end{equation}
On the other hand,
from Eq.~(\ref{eq:multiphase-matching-condition}),
we have
\begin{equation}
\phi_{m}=-2{\rm arccot}\left(\sqrt{1-\gamma^{2}}\tan\left(2\pi m/L\right)\right),
\label{eq:phi_m}
\end{equation}
where $\gamma=T_{1/L}^{-1}\left(1/\delta\right)$, $L=2l+1$, and $\delta$ meets the condition of Eq.~(\ref{eq:delta-condition}).
Then, substitute
$\delta$ and $\gamma$ into Eq.~(\ref{eq:phi_m}), we obtain
\begin{equation}
\frac{\cot^{2}\left(\frac{\phi_{m}}{2}\right)}{\lambda-\sin^{2}\left(\frac{\pi}{2L}\right)}=\frac{\tan^{2}\left(\frac{2m\pi}{L}\right)}{\cos^{2}\left(\frac{\pi}{2L}\right)}.
\label{eq:phi_m_equ}
\end{equation}
Note that, according to Eq.~(\ref{eq:m}), $4m\pi=\left(L\pm1\right)\pi$, then we have
\begin{equation}
\tan^{2}\left(\frac{4m\pi}{2L}\right)=\cot^{2}\left(\frac{\pi}{2L}\right).
\label{eq:trigonometry_equ}
\end{equation}
Therefore, based on Eqs.~(\ref{eq:phi_s_equ}), (\ref{eq:phi_m_equ})
and (\ref{eq:trigonometry_equ}), we can see that
\begin{equation}
\left|\phi_{m}\right|=\left|\phi_{{\rm s}}\right|\label{eq:phi_m_eq_phi_s}.
\end{equation}
Finally, due to $\phi_{\frac{l+1}{2}}>0$ and $\phi_{\frac{l}{2}}<0$, Eq.~(\ref{eq:relation-phi_m-phi_s}) holds.

\section{Conclusion \label{sec:Conclusion}}

In summary, we have studied the new application of the analytical multiphase
matching (AMPM) condition specially for the case of known $\lambda$, i.e., based on the AMPM condition, we designed a quantum search
algorithm with 100\% success probability for any given $\lambda\in(0,1)$.
We derived all maximum points (value of 100\%)
of the success probability after applying the analytical matched-multiphase Grover
operations $l$ times, and further obtained the available number of
iterations, phases, and other parameters, which ensure $\lambda$ just fall at a
certain maximum point.
Moreover, as an example, we experimentally verified the single-qubit and single-solution algorithm for all possible Oracles.
Experimental results agree well with the theoretical expectations,
confirming the feasibility of the proposed algorithm.
The number of
iterations of our algorithm is up to 1 more than the original Grover
algorithm, and achieves the optimal level of the existing exact quantum search algorithms.
In addition, we theoretically proved that in our exact algorithm
based on the AMPM condition there coincidentally exists a phase
of which the absolute value is exactly equal to that of the phase
in the exact algorithm based on the single-phase matching condition.
Our study confirms the usefulness of the AMPM condition in the case of known $\lambda$,
and also provides a guideline to understand the
mechanism and expand more applications of this condition.

\section*{Acknowledgements}

We thank He-Liang Huang, Jing-Yi Cui, Jia-Ji Li and Jie Lin, for helpful discussions. We also
acknowledge the use of IBM's Quantum Experience for this work. The
views expressed are those of the authors and do not reflect the official
policy or position of IBM or the IBM Quantum Experience team. This
work was supported by the National Natural Science Foundation of China
(Grants No. 11504430 and No. 61502526) and the National Basic Research
Program of China (Grant No. 2013CB338002).





\section*{References}

\bibliographystyle{elsarticle-num}

\end{document}